\documentclass[%
 aip,
 amsmath,amssymb,
preprint,%
floatfix
]{revtex4-1}

\usepackage{graphicx}
\usepackage{dcolumn}
\usepackage{bm}

\usepackage{xcolor}
\usepackage{lipsum}

\usepackage[utf8]{inputenc}
\usepackage[T1]{fontenc}
\usepackage{mathptmx}
\usepackage[font=footnotesize]{caption}
\usepackage[colorlinks=true,linkcolor=blue]{hyperref}%

\begin{document}

\preprint{}

\title[]{Generation of imprinted strain gradients for spintronics}

\author{G. Masciocchi}
\email{gmascioc@uni-mainz.de}
 \affiliation{Institute of Physics, Johannes Gutenberg University
Mainz, Staudingerweg 7, 55099 Mainz, Germany}
 \affiliation{Sensitec GmbH, Walter-Hallstein-Straße 24, 55130 Mainz, Germany}

\author{M. Fattouhi}%
\affiliation{ 
Universidad de Salamanca, Department of Applied Physics, E-37008 Salamanca, Spain 
}%

\author{E. Spetzler}
\affiliation{Institute for Materials Science, Kiel University, Kaiserstraße 2, 24143 Kiel, Germany}

\author{M.-A. Syskaki}
\affiliation{Institute of Physics, Johannes Gutenberg University
Mainz, Staudingerweg 7, 55099 Mainz, Germany}
\affiliation{Singulus Technologies AG, Hanauer Landstrasse 107, 63796 Kahl am Main, Germany}

\author{R. Lehndorff}%
\affiliation{ 
Allegro MicroSystems Germany GmbH, Vangerowstraße 18/1, 69115 Heidelberg, Germany 
}%

\author{E. Martinez}%
\affiliation{ 
Universidad de Salamanca, Department of Applied Physics, E-37008 Salamanca, Spain 
}%

\author{J. McCord}
\affiliation{Institute for Materials Science, Kiel University, Kaiserstraße 2, 24143 Kiel, Germany}
\affiliation{Kiel Nano, Surface and Interface Science (KiNSIS), Kaiserstraße 2, 24143 Kiel, Germany}

\author{L. Lopez-Diaz}%
\affiliation{ 
Universidad de Salamanca, Department of Applied Physics, E-37008 Salamanca, Spain 
}%

\author{ A. Kehlberger}
 \affiliation{Sensitec GmbH, Walter-Hallstein-Straße 24, 55130 Mainz, Germany}

\author{M. Kläui}
\email{klaeui@uni-mainz.de}
\affiliation{Institute of Physics, Johannes Gutenberg University
Mainz, Staudingerweg 7, 55099 Mainz, Germany}
\date{\today}

\begin{abstract}

In this work, we propose and evaluate an inexpensive and CMOS-compatible method to locally apply strain on a Si/SiOx substrate. Due to high growth temperatures and different thermal expansion coefficients, a SiN passivation layer exerts a compressive stress when deposited on a commercial silicon wafer. Removing selected areas of the passivation layer alters the strain on the  micrometer range, leading to changes in the local magnetic anisotropy of a magnetic material through magnetoelastic interactions. Using Kerr microscopy, we experimentally demonstrate how the magnetoelastic energy landscape, created by a pair of openings, in a magnetic nanowire enables the creation of pinning sites for in-plane vortex walls that propagate in a magnetic racetrack. We report substantial pinning fields up to 15 mT for device-relevant ferromagnetic materials with positive magnetostriction. We support our experimental results with finite element simulations for the induced strain, micromagnetic simulations and 1D model calculations using the realistic strain profile  to identify the depinning mechanism. All the observations above are due to the magnetoelastic energy contribution in the system, which creates local energy minima for the domain wall at the desired location. By controlling domain walls with strain, we realize the prototype of a true power-on magnetic sensor that can measure discrete magnetic fields or Oersted currents. This utilizes a technology that does not require piezoelectric substrates or high-resolution lithography, thus enabling wafer-level production.

\end{abstract}

\maketitle




One of the promising "Beyond COMS" technologies are nanomagnetic and spintronic devices due to their non-volatile nature, high operating speed, extremely low power consumption, and well explored routes to read and write data\cite{manipatruni2018beyond}. One example is nanomagnetic tracks, where information (stored in domain walls - DWs) is propagated and manipulated by dipolar interaction along soft ferromagnetic nanowires\cite{ono1999propagation,atkinson2003magnetic}. The manipulation of DWs has quite a long history and a turning point in this research area was the demonstration of a current-controlled magnetic DW shift register \cite{hayashi2008current,kumar2022domain} (racetrack memory). Since then, more work has been done on the development of DW-based memories\cite{franken2012shift}, logic devices\cite{luo2020current} and sensors\cite{zhang2018magnetoresistive,diegel2009new,khan2021magnetic,borie2017geometrically} or neuromorphic computing\cite{alamdar2021domain,yue2019brain}. However, feasibility of the fabrication process and compatibility with existing CMOS devices must be ensured before full technological realization is achieved. 

One of the key challenges with these devices is the control of DWs\cite{boulle2011current}, typically realized using geometric constraints (notches) \cite{omari2019toward, shiu2021depinning, al2019staggered} or the local manipulation of the magnetic anisotropy through strain \cite{fattouhi2022absence, masciocchi2021strain} using  magnetostrictive/piezoelectric systems \cite{lei2013strain, zhou2020voltage, hu2016fast, dean2011stress}. However, these approaches are not attractive for most sensor manufacturers due to high cost and complexity. Respectively, because high-resolution notches  and presence of  the multiferroic stack would require significant investments  in tools for high-resolution lithography and layer deposition. Also, the presence of voltages for piezoelectric actuation via metallic contacts increases design complexity and area usage. It is moreover difficult to realize an arbitrary shape of strain and strain gradients down to the micrometer range with piezoelectric substrates because it is technologically nontrivial to confine the electric fields\cite{barra2021effective}.

An alternative method of transferring strain to a thin film\cite{chu2009strain,doherty2020capping}, is the use of capping layers \cite{martin2009local} widely used in the semiconductor and photovoltaic industries because they provide protection from hostile environments.

In this work, we propose and experimentally demonstrate a low-cost and CMOS-compatible method to induce local strain on a Si/SiOx substrate by removing selected regions of the passivation layer. Arbitrary strain magnitudes and strain gradients can be realized by simply choosing the design of the removed part. The magnitude and profile of the strain are determined by combining anisotropy and stress measurements with finite elements simulations.  We experimentally demonstrate, using Kerr microscopy, that this local strain allows for domain wall pinning in a racetrack element. This is verified by micromagnetic simulations and 1D model calculations. Finally, to show the technological relevance of this method, we propose and verify a non-volatile magnetic peak-field sensor based on this technology. 


\begin{figure}[ht]
\centering\includegraphics[width=9cm]{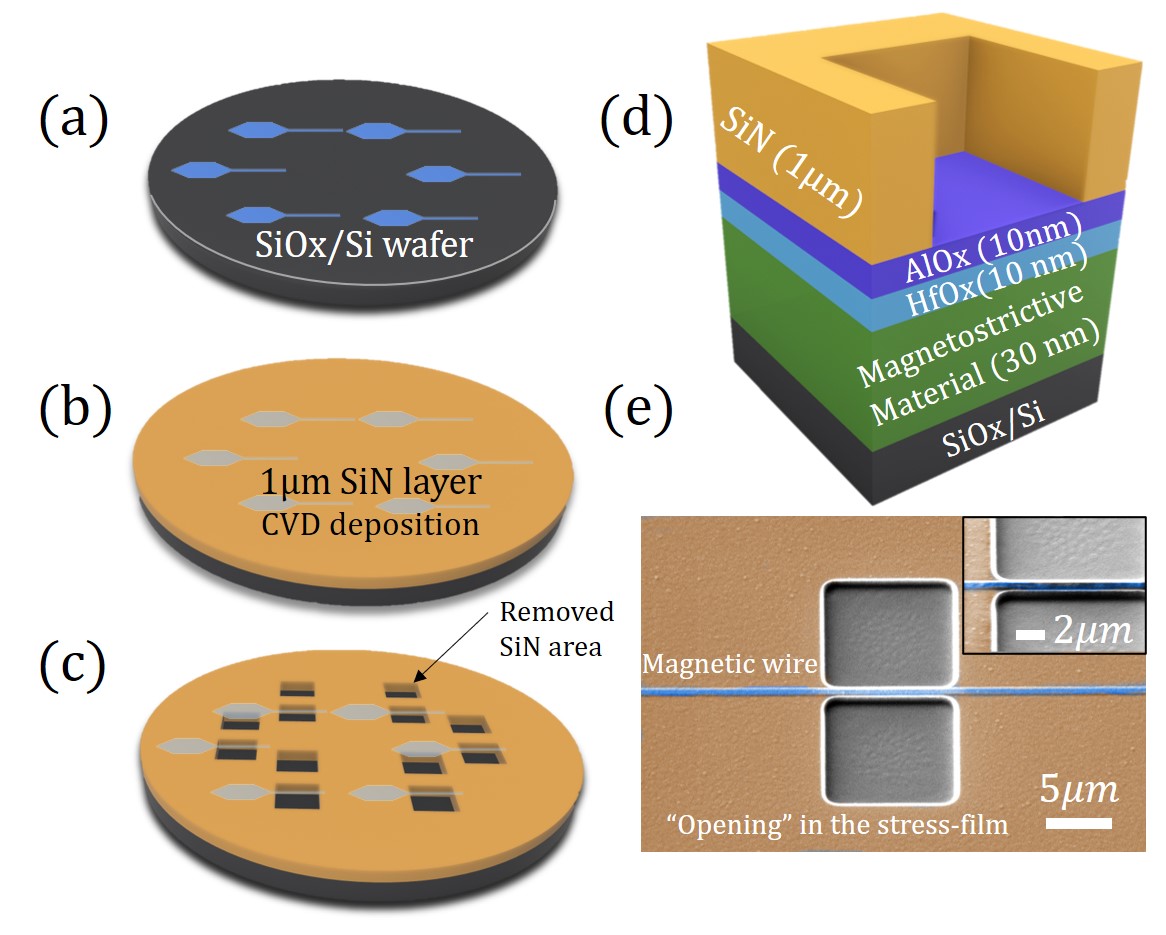}
\caption{\label{fig_1}  Racetracks structuring (a), SiN layer deposition (uniform strain generated) (b), and locally relieved strain after RIE (c). (d) section of the layers used in this work. (e) SEM image of a pair of openings in the SiN in the proximity of a magnetic nanowire. }
\end{figure}


Samples of AlOx(10 nm)/HfOx(10)/Co$_{70}$Fe$_{30}$(30) and AlOx(10 nm)/HfOx(10)/Co$_{40}$Fe$_{40}$B$_{20}$(30) were prepared by DC magnetron sputtering using a Singulus Rotaris system on a SiOx(1.5 µm)/Si(625 µm) substrate. Using optical lithography and etching, nanowires were fabricated with a variable width - between 800 and 500 nm - and a length of 70 µm. A reservoir at the left end allows for DWs injection at lower fields.
After the first lithography step, the wafer was covered with a 1 µm thick SiN layer using plasma-enhanced chemical vapor deposition (PECVD) at a temperature of 250$^\circ$C. The residual stress on the wafer is quantified using a standard wafer bow measurement. A second optical lithography step is used in combination with reactive ion etching (RIE) to remove selected areas (up to $20\times20$ µm$^2$) of the SiN layer (openings) without damaging the magnetic layer, as shown in Figs.\ref{fig_1} (a) - (d) while the wafer surface is still largely covered.
The hysteresis loops of the thin films are measured using a BH-looper with a setup to measure magnetostriction. The magneto-optic Kerr effect (MOKE) was used to image the magnetization state in the devices\cite{mccord2015progress,soldatov2017selective}.

To understand the origin of the intrinsic stress in our system, one should consider the coefficients of thermal expansion of a film and a substrate, along with the high temperature during deposition.  If the thermal expansion coefficients are different, thermal stresses arise when the whole stack cools down to room temperature after deposition \cite{ohring1992materials}. Relaxation of this stress leads to a deformation, i.e. bending, of the wafer (Figs. \ref{fig_1}(a)-(b)) allowing for the residual  stress to be estimated\cite{marks2014characterization}. The measured in-plane (compressive) stress is planar and in our case has a magnitude of $-495(5)$ MPa. To create a non-uniform stress on the substrate surface, selected areas of the SiN are completely removed, creating some openings in the passivation layer (Fig. \ref{fig_1} (c)). The etching is monitored to stop the process at the AlOx/SiN interface, as shown in Fig. \ref{fig_1} (d), so that the integrity of the magnetic layer is preserved. An example of the final device is shown in a scanning electron microscopy image (SEM)  in Fig. \ref{fig_1} (e) for a pair of square apertures  $10\times10$ µm$^2$ in size. With a suitable lithography mask, arbitrary shapes, sizes, and spacing of the apertures can be realized with sub  µm resolution. In the example presented here, the openings are spaced 1 µm apart and the 800 nm wide magnetic track under the SiN layer shows no signs of damage caused by the etching process.

  \begin{figure*}[ht]
\centering\includegraphics[width=12cm]{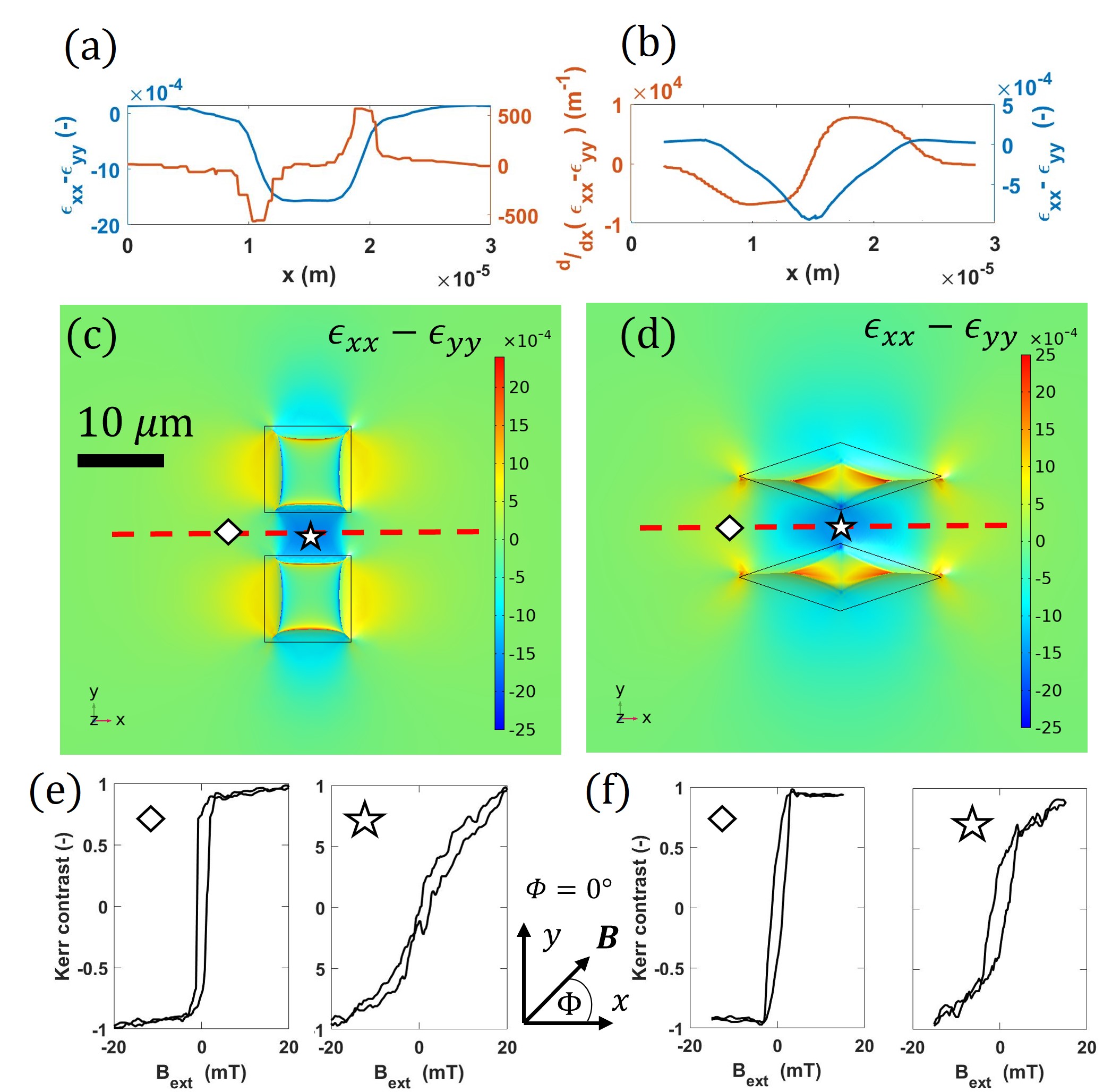}
\caption{\label{fig_2} (a) - (b) effective strain ($\epsilon_{xx}-\epsilon_{yy}$) and and strain gradient ($\frac{d}{dx}\left( \epsilon_{xx}-\epsilon_{yy} \right)$) for, respectively, a square and a triangular pair of opening obtained from FEM simulations. The $x$ axis refers to the red dashed line. (c) - (d) surface strain $\epsilon_{xx}-\epsilon_{yy}$ obtained with FEM simulations. (e) - (f) in-plane hysteresis loops obtained with Kerr microscopy on a full film of Co$_{70}$Fe$_{30}$ (30 nm) for, respectively, a square and a triangular pair of opening. The contrast was measured before the opening (diamond) and between them (star) according to the marker position. The magnetic field was applied along the direction $\Phi=0^\circ$. }
\end{figure*}

To determine the magnitude of stress relieved, finite-element-method (FEM) simulations were performed using the COMSOL Multiphysics® Structural Mechanics Module\cite{comsol}. More details about FEM simulations can be found in Section S1 of the supplementary material. To have a well defined strained region in racetrack type devices, it is convenient to consider a pair of openings - to be realized at each side of a magnetic nanowire. Figs. \ref{fig_2} (a) - (d) contain the computed values of the surface strain $\epsilon_{xx}-\epsilon_{yy}$ at the interface between SiN/SiOx for two different opening geometries. As shown in Figs. \ref{fig_2} (c) and (d), the effective surface strain $\epsilon_{xx}-\epsilon_{yy}$ is close to zero at a distance greater than 20 µm  from the etched areas and becomes non-uniform in their proximity. The geometry of the opening determines the strain profile. This can be seen in Figs. \ref{fig_2} (a) and (b) where the effective strain is plotted along the dashed line running between the two openings shown in Figs. \ref{fig_2} (c) and (d), respectively. For a square pair of openings (Fig. \ref{fig_2} (a)), the effective uniaxial strain profile $\epsilon_{xx}-\epsilon_{yy}$ is mostly flat and confined between them with strain gradient maxima (minima) at the exit (entrance) of the strained area. The strain reaches values of $\epsilon_{xx}-\epsilon_{yy}\simeq0.2\%$. For a diamond shaped pair of openings, the strain is again confined between the openings but its magnitude increases almost linearly towards the center. This time the strain gradient is mostly constant. 


To experimentally confirm the magnitude and sign of this local strain, we measured the magnetization curves of a unpatterned film of AlOx/HfOx/Co$_{70}$Fe$_{30}$ (30 nm) underneath the patterned SiN. The hysteresis loops were measured with field applied along $\Phi=0^\circ$ at different locations on the sample, selecting a region of interest of $5\times5$ µm$^2$ size within the field of view of the Kerr microscope. Full angular dependence of the anisotropy is reported in section
S3 of the supplementary information. Since Co$_{70}$Fe$_{30}$ has considerable magnetostriction ($\lambda_s\simeq 80\times10^{-6}$), the strain acting on the film is coupled to the magnetization via the magnetoelastic effect, as expressed in the anisotropy energy\cite{finizio2014magnetic}
  \begin{equation} \label{eq_strain_eanis}
K_{ME}=\frac{3}{2}\lambda_s Y \left( \epsilon_{xx}-\epsilon_{yy}\right),
\end{equation}
where $Y$ is the Young's modulus and $\lambda_s$ is the saturation magnetostriction. As done in previous works \cite{masciocchi2021strain,masciocchi2022control} the in-plane magnetoelastic anisotropy $K_{ME}$ can be locally estimated. Measuring hysteresis loops, where an in-plane field is applied along two perpendicular directions, can give us a direct measurement of the local strain\cite{mccord2004irregular,urs2014origin,thorarinsdottir2022finding}.

 In Figs. \ref{fig_2} (e) - (f) hysteresis loops of an unpatterned film, this time of SiN/AlOx/HfOx/Co$_{70}$Fe$_{30}$(30 nm), are shown. The openings geometry is the one of Figs. \ref{fig_2} (c) - (d), respectively.  Looking at Fig. \ref{fig_2} (e) we can compare the magnetization curve before (diamond) and between (star) the square openings. The anisotropy field increases, due to (uniaxial) magnetoelastic anisotropy. As Co$_{70}$Fe$_{30}$ has a positive magnetostriction, the increase in anisotropy field ($K_{ME}\simeq 8.9(2)
$ kJ/m$^3$) is caused by a negative (compressive) $\epsilon_{xx}-\epsilon_{yy}$ strain, in agreement with our FEM simulation. Using  Eq. \ref{eq_strain_eanis} and the values of magnetoelastic anisotropy difference we can estimate the strain  to be $\epsilon_{xx}-\epsilon_{yy}\simeq -0.05(1)\%$ for a square opening of this size. The same measurement can be performed for a diamond-shaped pair of openings and is reported in Fig. \ref{fig_2} (f). The calculated maximum strain difference for this case is $\epsilon_{xx}-\epsilon_{yy}\simeq -0.02(1)\%$.
 

The strain, created by removing specific areas of the SiN layer could be used as a mechanism to move, change direction, or stop a DW, a feature often needed in the device implementation\cite{masciocchi2021strain,diegel2009new}. Typical ways to do so relies on the modification of the DW energy making it a spatially variable quantity.
In analogy with the conventional field-driven case, the magnetoelastic field can be considered as a force that pushes the DW along the direction of
decreasing energy, i.e., increasing compressive strain if $\lambda_s>0$ for the in-plane-strain-gradient case. This force is proportional to the local gradient of the spatially variable
quantity \cite{wen2020ultralow, fattouhi2022absence,fattouhi2021electric}, and its effect is essentially that of an effective (magnetoelastic) field
  \begin{equation} \label{eq_strain_eff_field}
B_{ME}=-\frac{1}{M_s}\frac{d u_{ME}}{dx},
\end{equation}
 where $u_{ME}$ is the magnetoelastic DW energy per unit area.

\begin{figure}[ht]
\centering\includegraphics[width=8cm]{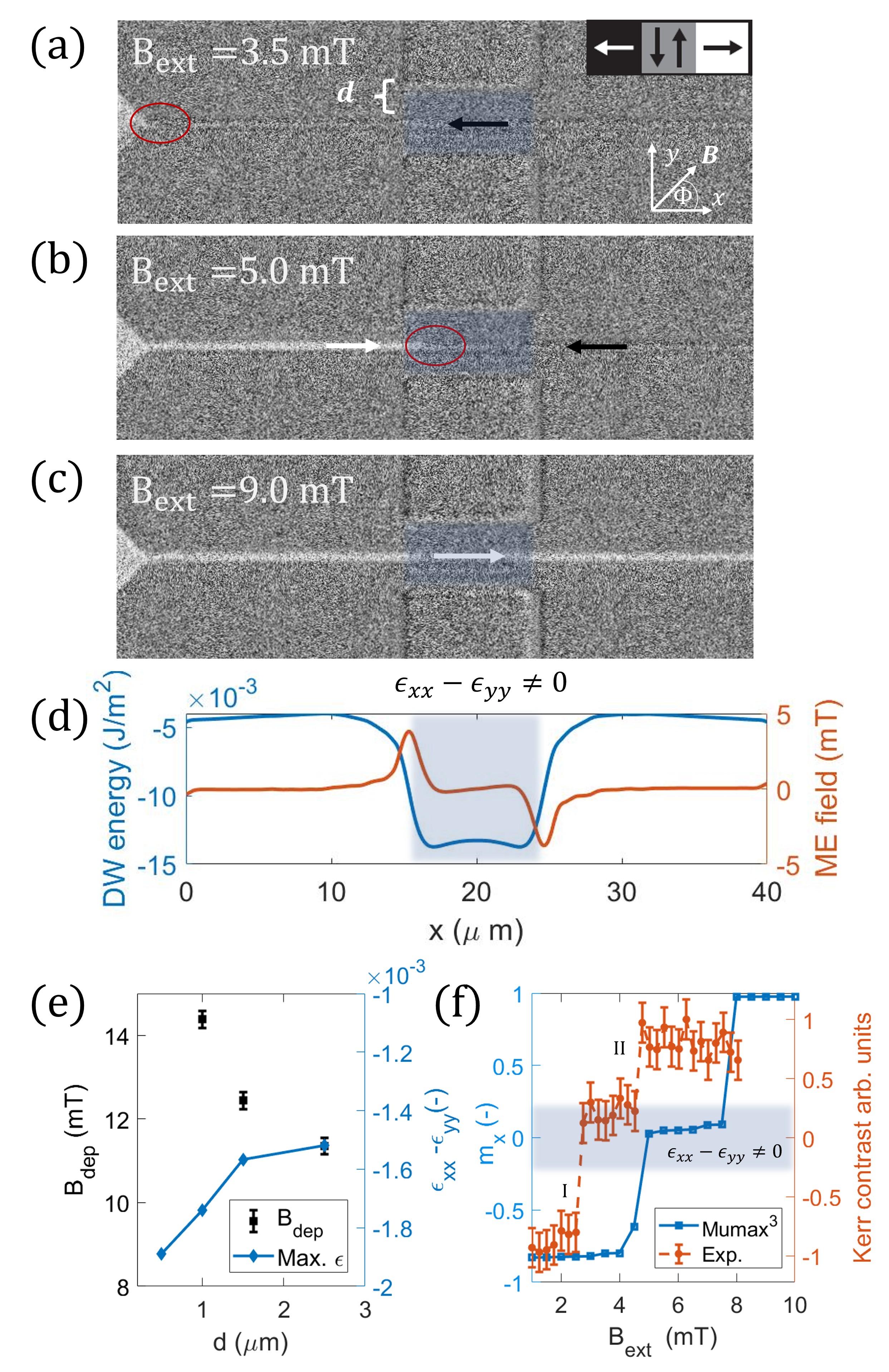}
\caption{\label{fig_3} MOKE images showing a DW (a) injected, (b) pinned in the strained area and (c) continuing propagation for larger magnetic fields. (d) 1D model calculations of energy profile  and the corresponding local magnetoelastic field for a vortex wall in the strain profile shown in Fig. \ref{fig_2}(a)  as a function of the DW position. (e) experimental values of depinning field (black squares) and maximum strain $\epsilon_{xx}-\epsilon_{yy}$ (blue diamonds) for different distances between a pair of square openings. The values consider a 500 nm wire of Co$_{70}$Fe$_{30}$ with 30 nm thickness.  (f) micromagnetic simulations magnetization (blue) and experimentally measured Kerr contrast (orange) for a 800 nm wide nanowire as function of applied field. The averaged wire magnetization along $x$ direction ($m_x$) is proportional to the DW position. }
\end{figure}

For this study, a  500 nm wide magnetic racetrack of Co$_{70}$Fe$_{30}$(30 nm) is considered together with a pair of square openings in the SiN 10$\times$10 µm$^2$ in size.   We use Kerr microscopy in transverse mode to image the magnetic state of the device, while in-plane magnetic field is applied parallel to the  wire along $x$. Figs. \ref{fig_3} (a)-(c) show the position of a DW along the magnetic racetrack as a function of the applied magnetic field. When the field is sufficiently large, the DW is injected from the reservoir (Fig. \ref{fig_3} (a)) into the magnetic wire. As can be seen in Fig. \ref{fig_3} (b), after injection the wall does not propagate until the end of the magnetic channel but  is pinned in the area between the SiN openings corresponding to the strained area. The corresponding surface strain was shown with a simulation in Figs. \ref{fig_2} (a) and (c). Only for larger magnetic fields the wall can continue to propagate to the other end of the magnetic channel, as captured by Fig. \ref{fig_3} (c).

 
For a Ni$_{81}$Fe$_{19}$ sample with nearly no magnetostriction  no DW pinning was found within the resolution of our measurement, supporting the idea of a strain-based pinning. We repeated the same measurement for devices with different distance $d$ between racetrack and openings. According to FEM simulations, the value of the (compressive) strain increases as the opening distance $d$ is reduced (blue diamonds in Fig. \ref{fig_3} (e)). As shown in Fig. \ref{fig_3} (e), the depinning field (black squares) increases  from 11.0(2) mT to 14.5(2) mT for a distance between the opening and the magnetic racetrack decreasing from 2.5 to 1 µm.  A larger depinning field $B_{dep}$ for smaller opening spacing confirms that  the magnetoelastic energy is indeed the dominant pinning cause in our system\cite{lei2013strain,franken2013voltage}.
 
 To support our experimental findings, we performed  micromagnetic simulations and 1D model calculations where the strain profile from FEM simulations was used. The results are summarized in Figs. \ref{fig_3} (d) and (f) and consider nanowires made of 30 nm thick Co$_{40}$Fe$_{40}$B$_{20}$. For more details about the micromagnetic simulations and the 1D analytical model, see section S2 of the supplementary information. Fig. \ref{fig_3} (d) shows the DW energy per unit area and the corresponding magnetoelastic field as a function of the DW position for a nanowire $w=500$ nm wide, calculated considering the strain profile shown in Fig. \ref{fig_2} (a) and a rigid profile for the DW.  Comparing Fig. \ref{fig_3} (d) with Fig. \ref{fig_3} (b), it is clear that the point where the DW sits is the minimum of DW energy. At the sides of the pinning site, the effective magnetoelastic field - proportional to $\frac{d}{dx}\left( \epsilon_{xx}-\epsilon_{yy} \right)$ according to Eq. \ref{eq_strain_eff_field} - is non-zero, and opposite to the applied external field. This equivalent force prevents the DW to move forward unless the external applied field is increased. 
 
\begin{figure}[ht]
\centering\includegraphics[width=8cm]{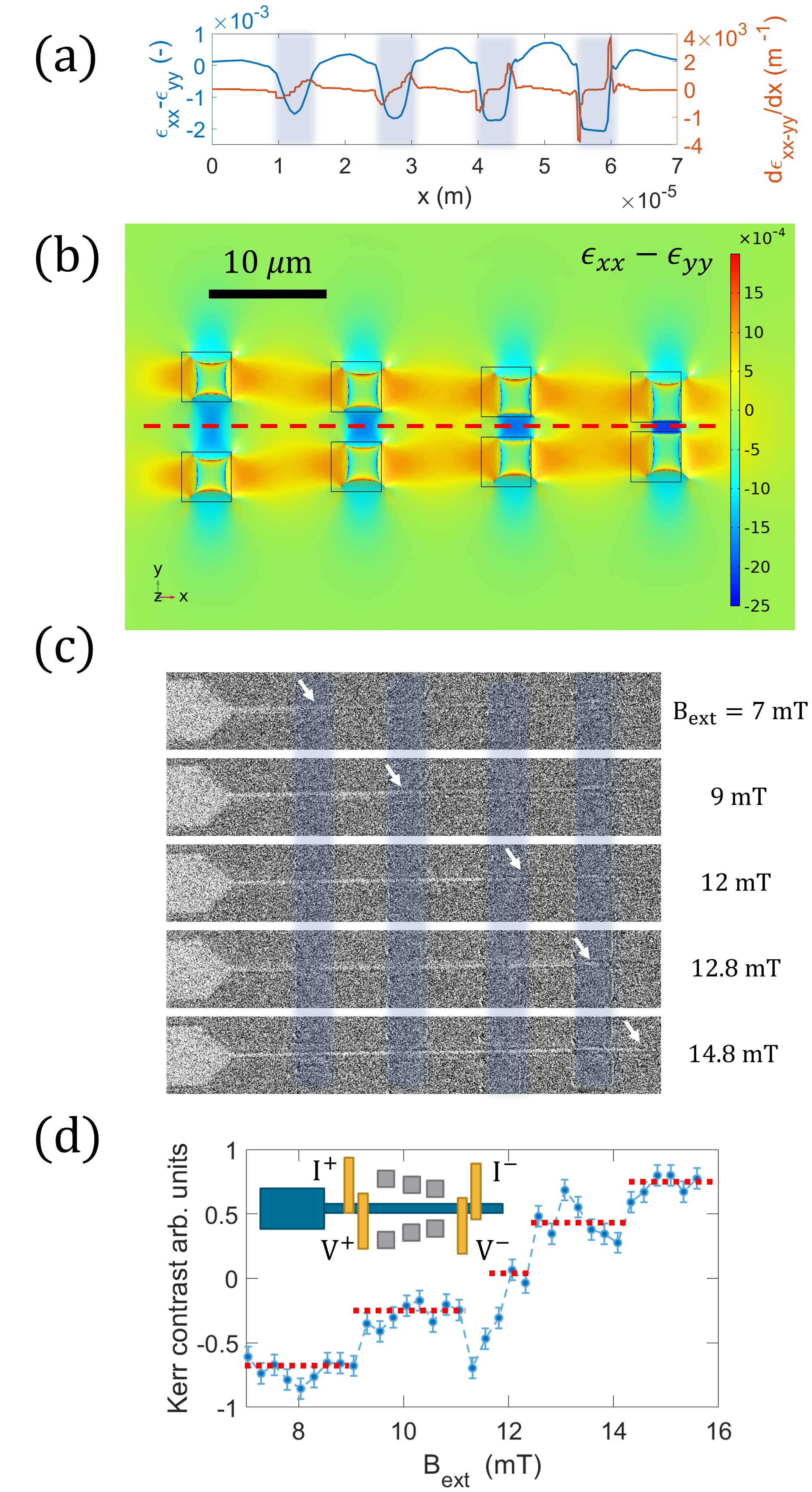}
\caption{\label{fig_4} (a) surface strain and strain gradient along the racetrack path, that is highlighted with a dashed line in (b) calculated with FEM. (c) MOKE images of the DW position (white arrow) in the device for different magnetic field steps. (d) magnetic contrast along the racetrack obtained with Kerr microscopy.  The sample is made of Co$_{70}$Fe$_{30}$ and the width of the wire is 500 nm. The inset shows a possible contacts configuration for resistivity measurements of the DW position.}
\end{figure}

  For the Mumax\cite{vansteenkiste2014design} micromagnetic simulations a wire  of 800 nm width has been considered. The magnetization has been initialized in the system with a DW on the left side of the strained area and then a magnetic field has been applied. Multiple dynamic simulations have been performed at different values of external magnetic field, and the results are summarized in Fig. \ref{fig_3} (f). As can be seen, the averaged magnetization along the $x$ direction (proportional  to the DW position) coincides with  the strained area (state I) for external fields $B_{ext}<B_{dep}$. When the applied magnetic field is increased above $B_{dep}$, the domain wall is free to propagate and reaches the right end of the wire (state II). For comparison, the experimental values for the DW position as a function of $B_{ext}$ is reported in Fig. \ref{fig_4} (f) for a 800 nm width Co$_{40}$Fe$_{40}$B$_{20}$ wire. The pinning position (where $\epsilon_{xx}-\epsilon_{yy}\neq 0$) coincides and discrepancies between the simulations and experiments for the value of $B_{dep}$ can be due to thermally activated depinning events that are not fully captured by micromagnetic simulations.




The ability to adjust the maximum value of the strain, and thus the value of the depinning field, by changing the aperture design - as shown in Fig. \ref{fig_3} (e) - allows for the realization of a non-volatile magnetic field sensor capable of detecting discrete values of magnetic fields or current peaks from wires or coils in the sensor proximity. Previous work \cite{al2019staggered}, suggested similar concepts, however, the one proposed here does not require sub 100 nm lithography resolution for the notches. 

The conceptualization of the peak-field sensor is presented in Fig. \ref{fig_4}. The device comprises of a magnetic nanowire for DWs propagation with a number of pinning sites along it.  As shown in Figs. \ref{fig_4} (a)  and (b) if the spacing between the SiN openings - acting as pinning sites  - decreases,  the strain magnitude is increased progressively. According to Fig. \ref{fig_3} (e), the depinning field $B_{dep}$ will increase going from left to right. The device considered here presents four pair of openings and is therefore able to identify four discrete levels of external magnetic fields.

The proof of concept of the realized device is presented in Figs. \ref{fig_4} (c) and (d). We consider, in this case, a magnetic wire of 500 nm width and realized with a Co$_{70}$Fe$_{30}$ magnetic layer. The shape of the openings is 5$\times$5 µm$^2$ and the distance between the pairs is, in order 5, 3, 2 and 1 µm, as shown in Fig. \ref{fig_4} (b).

The device is initialized with large, negative magnetic field in the $x$ direction creating a uniform magnetic state. After that, positive magnetic filed is gradually increased while the magnetic state in the wire is monitored using Kerr microscopy. The magnetic contrast in Fig. \ref{fig_4} (d) shows how the magnetization has well defined discrete levels as $B_{ext}$ is increased.  This occurs because a DW propagating into the nanowire occupies only discrete positions along $x$, as shown in Fig. \ref{fig_4} (c) in the strained area between the openings. The position of the DW in the magnetic channel (output) will indicate the maximum field (input) that the device has seen after initialization. The magnetic state has been measured in Fig. \ref{fig_4} (d) with optical methods, however, electrical readout of the DW position is possible using, e.g.,  Giant Magnetoresistive effect (GMR)\cite{diegel2009new} and two electrical contacts at the extremity of the magnetic channel. This sensing solution is particularly suitable for hardly accessible measurement environments and energy efficient devices as electrical power is required only for readout and initialization.

In summary, in this work we propose an validate a method for generating a local strain on a rigid substrate that is compatible with standard CMOS technologies. The intrinsic stress that occurs at the substrate/layer interface during SiN deposition can be modified when selected regions of the passivation layer are removed by etching. The strain is only modified near the removed material, as shown by FEM simulations. Using in-situ measurements of the magnetoelastic anisotropy, we experimentally determine the magnitude of the uniaxial strain up to $0.05(1)\%$. The magnitude and the gradient of the in-plane strain can be tuned depending on the geometry and position of the openings in the stress-generating layer. We validate the use of the above-mentioned strain gradients for the manipulation of magnetic domain walls in spintronic devices by exploiting magnetoelastic coupling in magnetostrictive materials. Using Kerr microscopy, we experimentally show how the magnetoelastic energy landscape enables the creation of engineered pinning sites which represent local energy minima for in-plane vortex walls. We report substantial pinning fields of up to 15 mT and support our experimental findings with micromagnetic simulations and 1D model calculations using a realistic strain profile. This provides the opportunity to realize an alternative generation of DW-based devices with technology compatible with wafer-level production, and an example of a discrete magnetic field or current sensor using imprinted strain gradients is demonstrated.

\section*{Supplementary Material}
See supplementary material for details about the material parameters used, the finite-element-method  and micromagnetic simulations and the anisotropy measurements.   

\begin{acknowledgments}
 This project has received funding from the European Union’s Horizon 2020 research and innovation program  under  the  Marie  Skłodowska-Curie  Grant  Agreement  No  860060  “Magnetism  and  the effect of Electric Field” (MagnEFi), the Deutsche Forschungsgemeinschaft (DFG, German Research Foundation) - TRR 173 - 268565370 (project A01 and B02),  the DFG funded collaborative research center (CRC)1261 / project A10  and the Austrian Research Promotion Agency (FFG). The authors acknowledge support by the chip production facilities of Sensitec GmbH (Mainz, DE), where part of this work was carried out and the Max-Planck Graduate Centre with Johannes Gutenberg University.
\end{acknowledgments}

\section*{Author Declarations}
\subsection*{Conflict of interest }
The authors have no conflicts to disclose.

\section*{Data Sharing Policy }
The data that support the findings of this study are available from the corresponding author upon reasonable request.





\nocite{*}
\bibliography{bibliography}

\newpage



\end{document}


\preprint{}

\title[\textbf{Suppl. material} - Generation of imprinted strain gradients for spintronics]{Supplementary material -  Generation of imprinted strain gradients for spintronics}

\date{\today}

\maketitle

\subsection*{\textbf{S1} - COMSOL simulations }

\begin{figure*}[h!]
\centering\includegraphics[width=14cm]{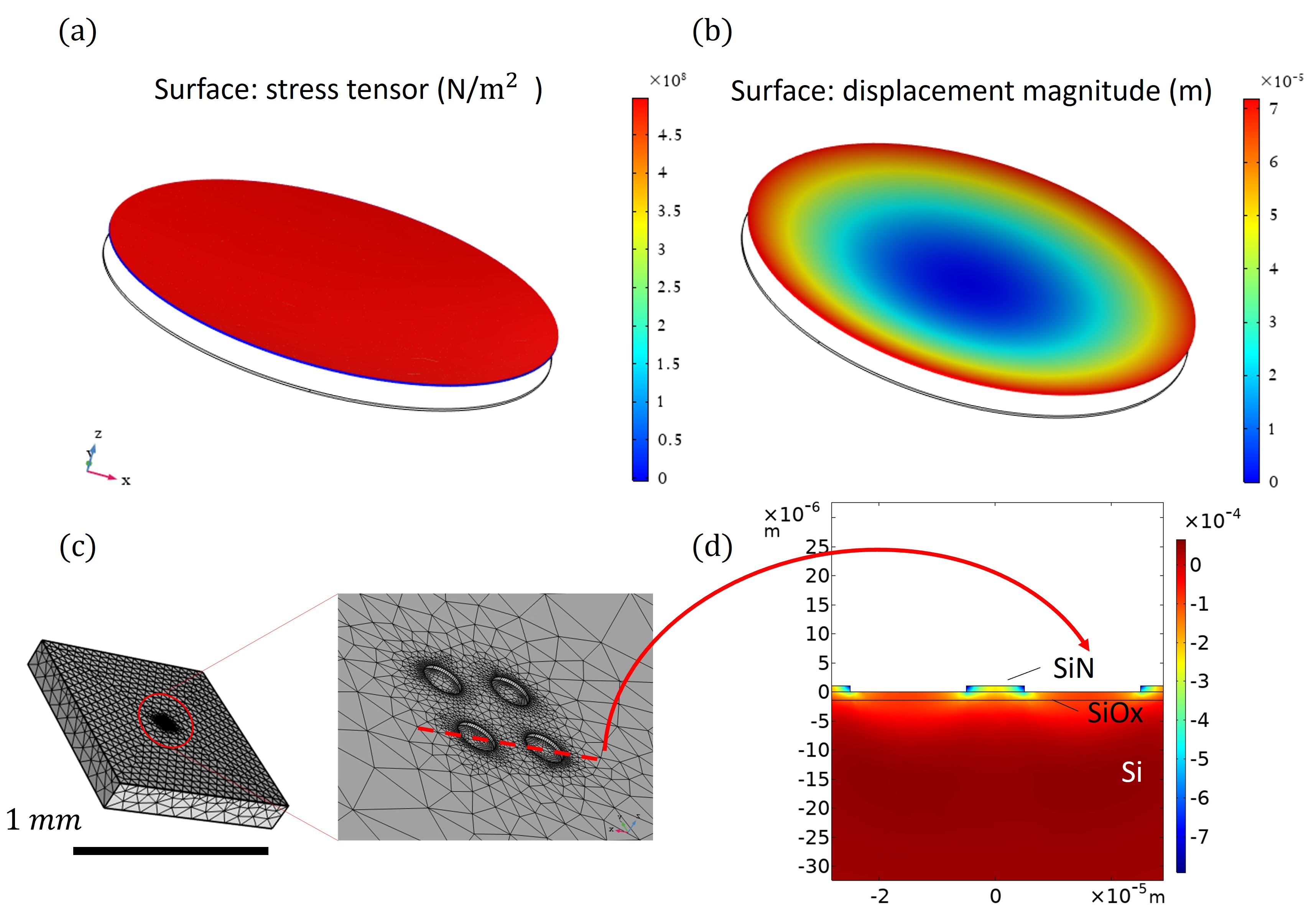}
\caption{\label{fig_S01}Details of the FEM simulations performed to obtain the stress/strain distribution. Stress tensor on a Si/SiOx wafer simulated using COMSOL Multiphysics after deposition of a tensile SiN over-layer. (a) stress tensor on the surface of the wafer which is uniform. (b) surface displacement of a 5" wafer. (c) discretization and (d) x component of the strain tensor simulated in the proximity of a removed compressive SiN layer. }
\end{figure*}

To simulate the stress and strain profile a numerical model with finite-element-method (FEM) discretization was built in COMSOL multyphysics software. The built in "Structural Mechanics" module and stationary study were used.  The simulation computes the relaxed state of the system after initial stress (extracted form wafer-bow measurements) is imposed. Figs. S\ref{fig_S01} (a)-(b) shows an example of the uniform stress induced on a 5" wafer due to the SiN layer deposition. 
In Fig. S\ref{fig_S01} (c)-(d)  some examples of the simulation output on a 1$\times$1 mm$^2$ sample in the presence of openings in the SiN are shown. In Fig. S\ref{fig_S01} (d) it is possible to see how the non-uniformity of the  stress created by the opening propagates for several hundreds of nm from the interface between SiN and SiOx.

\begin{table}[h!]
    \centering
    \begin{tabular}{||c c c c c||} 
 \hline
    Material & $\rho$ (Kg/m$^3$) & $Y$ (GPa)  &  $\nu$ (-) & Thickness (m) \\ [0.5ex] 
 \hline\hline
 Si (Poly-silicon)\cite{parameters_Si} & 3.32 $\times10^{3}$ &	169 &	0.22 & 625$\times10^{-6}$\\
  \hline
 SiOx\cite{parameters_SiOx} & 2.2 $\times10^{3}$ &	70 &	0.17 &	1.5$\times10^{-6}$  \\ 
 \hline
 SiN\cite{parameters_SiN} &  3.1$\times10^{3}$  &	250 &	0.23 &	1$\times10^{-6}$ \\ 
 \hline
\end{tabular}
\caption{Parameters used for the FEM simulations. Here, $\rho$ is the material density, $Y$ is the Young's modulus and $\nu$ is the Poisson's ratio. The values are taken from literature. }
\label{tab_comsol}
\end{table}

 The initial in-plane stress imposed on the sample is -500 MPa (compressive), as measured experimentally. To calculate the variation in the surface strain caused by the etched areas in the SiN, the size of the system was reduced to 1$\times$1 mm$^2$ to have reasonable computational time. All the relevant material parameters used are reported in Table S\ref{tab_comsol}.


\subsection*{ \textbf{S2} - Micromagnetic simulations and 1D Model}

The simulated system is nanowire of 30 nm of thickness. System size are 40$\times$0.8  µm$^2$ and the cell size of 5 nm is below the exchange lenght of CoFeB.

The material parameters used\cite{wang2017,yu2015,o_handley} are the one for Co$_{40}$Fe$_{40}$B$_{20}$: the elastic constants are $C_{11}$=280 GPa, $C_{12}$=140 GPa, $C_{44}$=75.5 GPa,  the magnetostriction is $\lambda_s$=2.9$\times10^{-5}$, saturation magnetization $M_s$=1$\times10^{6}$ A/m, the exchange constant $A_{ex}$=15$\times10^{-12}$ and the damping is set to $\alpha$=0.015. Disorder in the system was introduced by varying  the material parameters $A_{ex}$ and $M_s$ of 5\% over 20 nm. A realistic edge roughness of 30 nm was considered and the temperature of the simulations is 300 K.  The applied strain is extracted from COMSOL simulations for the corresponding SiN opening geometry.

 \begin{figure*}[h!]
\centering\includegraphics[width=14cm]{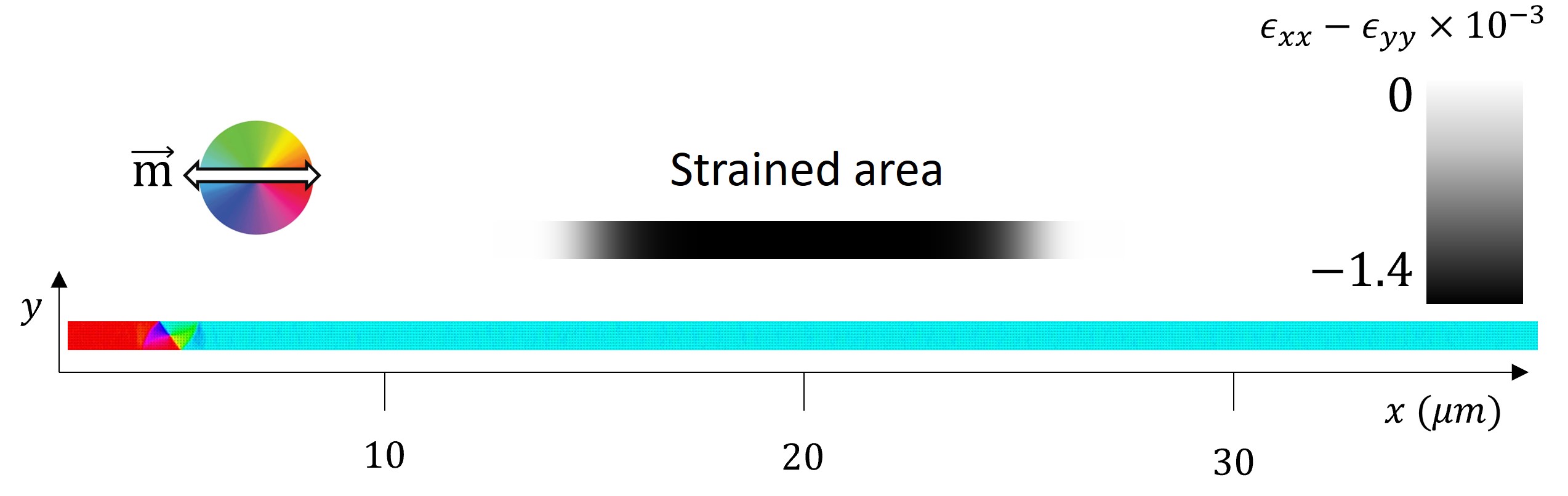}
\caption{\label{fig_S02}Details of the Mumax3 simulations performed. The used strain profile, comes form a pair of squared openings. The snapshot of the magnetization was taken just before applying the external field (along $x$).  }
\end{figure*}

After initializing the magnetization with a vortex wall on the left side of the wire, an external magnetic field is applied to the system and moves the DW in the direction of the strained area as shown in Fig. S\ref{fig_S02}. Multiple dynamic simulations have been performed for each value of the applied field that is increased until the domain wall continues propagation. The time evolution was monitored  for 200 ns, a time long enough for the DW to reach the pinning position in the field range used. Micromagnetic simulations reproduce the DW pinning position and the relative strength of their corresponding depinning fields but with a difference (2.5 mT) in their values if compared with the experimental case. This mismatch is mostly given by the fact that in our simulations at T=300 K the field is swept in steps of 0.5 mT every 200 ns, a time scale much shorter than the experimental one. Simulations on a time scale similar to the experiment would yield lower depinning fields, since thermally activated depinning would play a much more important role, but they are unrealizable. For this reason, a quantitative agreement  is beyond the scope of the manuscript.

In addition to micromagnetic simulations, 1D model calculations have been performed to extract the domain wall energy as a function of the DW position in the strain profile created by the etched areas in the SiN\cite{martiney2006domain}. The 1D strain profile, $\epsilon_{xx}$ and $\epsilon_{yy}$ is calculated form FEM simulations. Considering that the magnetoelastic energy is given by\cite{forrest1964uniaxial} 
 \begin{equation} \tag{S.1}\label{eq_magel_en}
u_{ME}=B_1\left(\epsilon_{xx}m_x^2+\epsilon_{yy}m_y^2\right)
\end{equation}
the DW energy (per unit area) as a function of the DW position, $x_0$, can be calculated by convolution with the DW profile in the 40$\times$0.5  µm$^2$ nanowire
 \begin{equation} \tag{S.2}\label{integral_magel_en}
U_{DW}(x_0)=B_1\int_{0}^{40\mu m} \left[\epsilon_{xx}m_x^2(x-x_0)+\epsilon_{yy}m_y^2(x-x_0)\right] dx,
\end{equation}
where $B_1$ is the magnetoelastic constant, and $m_x$, $m_y$ are the x and y component of the magnetization, respectively. The magneto-elastc field  is then trivially the first derivative of the magneto-elastic energy according to\cite{martinez_2006_PRB}
 \begin{equation} \tag{S.3}\label{DW_ME_field}
B_{ME}=-\frac{1}{M_s}\frac{d u_{ME} }{dx}. 
\end{equation}

The profile of the wall considered is
 \begin{equation} \tag{S.4}\label{DW_mx}
m_x(x-x_0)= \cos \left[ \tan^{-1} \left[ exp\left( -\frac{x-x_0}{w/2} \right) \right] \right] 
\end{equation}
for the x and 
 \begin{equation} \tag{S.5}\label{DW_my}
m_y(x-x_0)= \sin \left[ \tan^{-1} \left[ exp\left( -\frac{x-x_0}{w/2} \right) \right] \right] 
\end{equation}
for the y component of magnetization, respectively,  where  $w$ is the wire width.

\begin{figure*}[h!]
\centering\includegraphics[width=14cm]{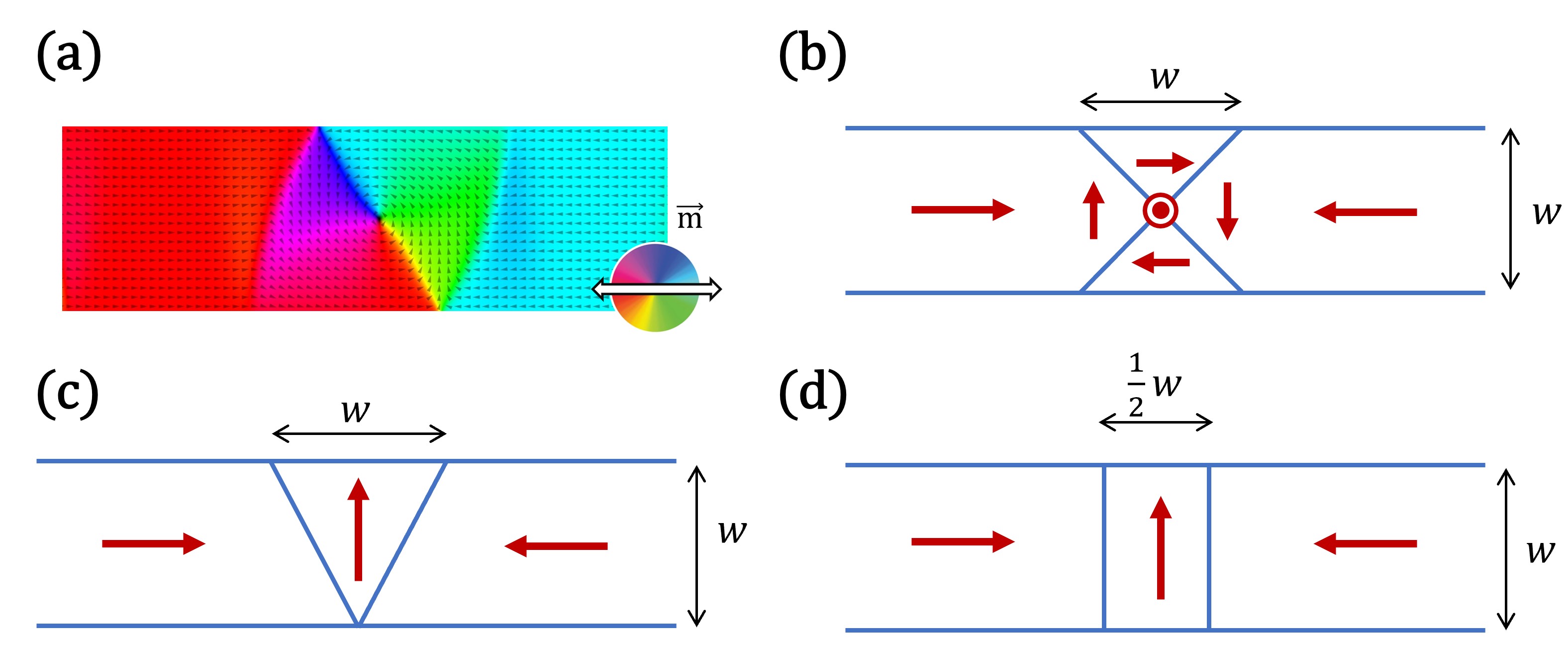}
\caption{\label{fig_S03} Simulated profile of an in-plane vortex wall in a wire whose width is $w$ (a) and schematic representation of it (b). (c) triangular wall of width $w$ and (d) transverse wall of width $\frac{1}{2}w$. }
\end{figure*}

The profile for the DW used is that of a 1D domain wall of width $\delta=w/2$. The crude approximation is well justified  on the following grounds:
firstly, a vortex DW extends over an area $\simeq w^2$ (Figs. \ref{fig_S03} (a) and (b)), so the effective width is $\simeq w$. Secondly, the magnetoelastic energy is only sensitive to the net $m_x$ and $m_y$ components of magnetization. Consequently, except for the vortex core (which occupies a very tiny area) a vortex wall and a (triangular) transverse wall of the same width (Fig. \ref{fig_S03} (c)) have the same net $m_x$ and $m_y$. Eventually, following the same argument, a triangular transverse wall of width $w$ and a 1D transverse wall of of width $w/2$ (Fig. \ref{fig_S03} (d)) also have the same net $m_x$ and $m_y$. In this model, it is assumed that exchange and magnetostatic energies do not change as a function of the DW position. In other words, we are assuming that the DW is not deformed as it moves through the strain profile.

In this way, the profile of the DW energy (per unit area) as a function of the DW position can be calculated. Additionally, the magnetoelastic field acting on the DW is straightforwardly obtained by deriving the DW energy profile in Eq. \ref{integral_magel_en}.
For the calculations, the same parameters used in Mumax simulations are considered. Both the 1D model and the micromagnetic simulations are qualitatively reproducing the observations and a quantitative comparison is beyond the scope of this experimental study.

\begin{table}[h!]
    \centering
    \begin{tabular}{||c c c c||} 
 \hline
    Material & $M_{s}$  (T)  & $\lambda_s$ x$10^{-6}$ & $Y$  ($GPa$) \\ [0.5ex] 
 \hline\hline
 $Co_{40}Fe_{40}B_{20}$ & 1.40 &	30 & 187\\
  \hline

 $CoFe_{30}$ &  2.0 &	80  & 180 \\ 
  
 \hline
\end{tabular}
\caption{Parameters from literature\cite{cullity2011introduction,bozorth1993ferromagnetism,peng2016fast} of the magnetic materials after deposition and used for the estimation of the magnetoelastic anisotropy. Here, $M_s$ is the saturation magnetization, $\lambda_s$ is the saturation magnetostriction and $Y$ is the Young's modulus. }
\label{tab_material_calc_film}
\end{table}


\subsection*{ \textbf{S3} - Strain characterization - local anisotropy measurements}

\begin{figure*}[ht]
\centering\includegraphics[width=12cm]{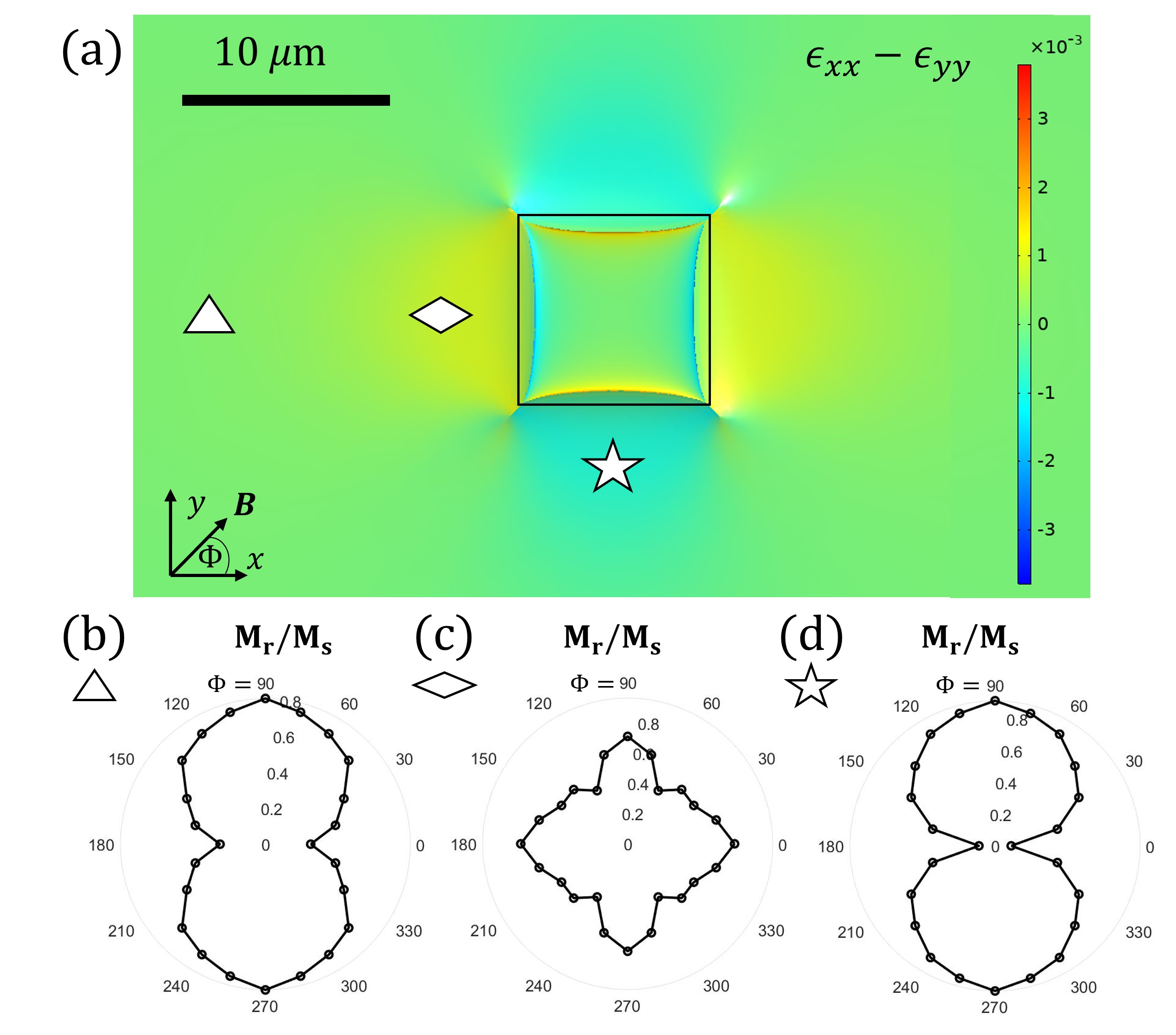}
\caption{\label{fig_S4} (a)  surface strain ($\epsilon_{xx}-\epsilon_{yy}$) for a single square-shaped  opening obtained from FEM simulations.  (b) - (d) angular plot of the normalized remanent magnetization $M_r/M_s$ as a function of the in-plane angle $\Phi$ of the magnetic field measured with Kerr microscopy on a full film of Co$_{40}$Fe$_{40}$B$_{40}$ (30 nm) where a single opening in the SiN is present. The measurements were taken in different locations around the etched area in the passivation layer as indicated by the markers. Namely, an area far from the opening (b), on the left side where $\epsilon_{xx}-\epsilon_{yy}>0$ (c) and on the lower side where $\epsilon_{xx}-\epsilon_{yy}<0$ (d) are considered. }
\end{figure*}

In this section, additional stress simulations and anisotropy measurements are reported for the case of a single square shaped opening in the SiN overlayer. Fig. S\ref{fig_S4} (a) contains the calculated values of the surface strain $\epsilon_{xx}-\epsilon_{yy}$ at the interface between SiN/SiOx. Fig. S\ref{fig_S4} (a) considers the simplest case of a single square aperture of $10\times10$  µm$^2$ in size. As shown in Fig. S\ref{fig_S4} (a), the surface strain is constant at a distance greater than $20$  µm from the etched regions and $\epsilon_{xx}-\epsilon_{yy}$ approaches zero, as expected from the residual stress measurements. However, near the openings (shown as black lines), the strain profile changes and reaches values of $\epsilon_{xx}-\epsilon_{yy}\simeq0.2\%$. According to the symmetry of the system, the $\epsilon_{xx}-\epsilon_{yy}$ strain is positive (tensile) to the right and left of the opening, while it is negative (compressive) above and below the opening itself.

To have a full understanding of the modification of the local anisotropy in a magnetostrictive film caused by an external strain, it is convenient to measure a full angular dependence of the magnetization curves as a function of the in-plane angle. Figs. S\ref{fig_S4} (b) - (d) show the angular dependence of the normalized remanent magnetization $M_r/M_s$ in three different position with respect to a square-shaped opening in the SiN. The values are plotted as a function of the in-plane angle of the applied magnetic field $\Phi$. This type of plot allows one to identify the direction of the effective in-plane magnetic anisotropy $K_{eff}$ in the material\cite{thorarinsdottir2022finding}. In Fig. S\ref{fig_S4} (b) the measurement of $M_r/M_s$ in a location 20  µm to the left of the opening (triangle in Fig. S\ref{fig_S4} (a)) is reported. A uniaxial magnetic anisotropy with easy axis along $\Phi=90^\circ$ is present and  $M_r/M_s(\Phi=0^\circ)\simeq0.2$ along the hard axis. According to COMSOL simulations, in this area the effective uniaxial strain along $x$ is close to zero. Therefore the observed anisotropy is presumably deposition-induced rather than strain-induced\cite{robert2000handley}. In regions of the sample in close proximity to the opening - where $\epsilon_{xx}-\epsilon_{yy}\neq0$ - the measured film magnetic anisotropy is altered with respect to the case in Fig. S\ref{fig_S4} (b).  Figs. S\ref{fig_S4} (c) and (d) considers area where the local  uniaxial strain along $x$  is tensile (diamond) or compressive (star), respectively. In Fig. S\ref{fig_S4} (c) the angular plot of $M_r/M_s$ indicates the presence of two uniaxial magnetic anisotropies with easy axis oriented along 0 (dominant) and 90$^\circ$. A first easy axis along $90^\circ$ is deposition induced. In addition to that, a magnetic easy axis along $0^\circ$ is induced by the magnetoelastic anisotropy. Given that Co$_{40}$Fe$_{40}$B$_{20}$ has a positive magnetostriction, the strain induced easy axis will be along the direction of the tensile uniaxial strain, which is along $x$ ($\Phi=0^\circ$) in this case. This is in agreement with our COMSOL simulations that give $\epsilon_{xx}-\epsilon_{yy}=+0.15\%$ in the area marked by a diamond in Fig. S\ref{fig_S4} (a). The same measurement for $M_r/M_s$ has been performed in an area where the effective strain along $x$ is compressive and is reported in Fig. S\ref{fig_S4} (d). In this region (marked by a star in Fig. S\ref{fig_S4} (a)) the strain-induced easy axis is expected to be along $y$ ($\Phi=90^\circ$) i.e. along the same direction of the deposition-induced anisotropy. Accordingly, the effective anisotropy is increased with $M_r/M_s(\Phi=0^\circ)\simeq0$ along the hard axis of magnetization.

\newpage
